\documentclass{article}[12pt]

\ifx\pdfoutput\undefined       \usepackage[dvips]{graphicx}      \else
\usepackage[pdftex]{graphicx} \pdfcompresslevel=9 \fi

\usepackage{epstopdf} 
\usepackage{subfigure}
\usepackage{graphicx}
\usepackage{textcomp}
\usepackage[usenames,dvipsnames]{color}
\usepackage{amsmath}
\usepackage{slashed}

\def\gray{{$\gamma$-ray\ }}

\begin{document}


\begin{center}
{\Large{\bf TESTING LORENTZ SYMMETRY USING}} 
\vspace{12pt}
{\Large{\bf HIGH ENERGY ASTROPHYSICS OBSERVATIONS}

\vspace{24pt}



{\bf Floyd W. Stecker} \\
{NASA Goddard Space Flight Center} \\
{Greenbelt, MD 20771 U.S.A.}}
\end{center}






\vspace{24pt}
\section*{Abstract}
We discuss some of the tests of Lorentz symmetry made possible by astrophysical
observations of ultrahigh energy cosmic rays, $\gamma$-rays, and neutrinos.
These are among the most sensitive tests of Lorentz invariance violation because they are
the highest energy phenomena known to man.

\vspace{24pt}

\newpage

\tableofcontents


\newpage

\section{Introduction}
One of the main motivations for testing Lorentz symmetry comes from the search for an answer to one of the most fundamental problems of modern physics: how to reconcile general relativity with quantum physics.~This problem is particularly acute on the length scale where the effect of quantum fluctuations on spacetime cannot be ignored.~This scale, known as the Planck length,
is given by \mbox{$\lambda_{Pl} = \sqrt{G\hbar/c^3} \sim 10^{-35}$ m~\cite{pl06}}. It corresponds to the very large energy scale, $E_{Pl} = M_{Pl}c^2 = 1.2 \times 10^{19}$~GeV, where $M_{Pl} = \sqrt{\hbar c/G}$.~(Hereafter, we use the conventions for the low energy speed of light $c = 1$ and $\hbar = 1$.)~The open question of constructing a framework for a quantum theory of gravity is intimately related to the question of Lorentz invariance violation~\cite{ko89,ta14}.

Lorentz symmetry implies scale-free spacetime. The group of Lorentz transformations is unbounded in energy (boost) space.~Owing to the uncertainty principle, the higher the particle energy attained, the smaller the scale of physics that can be probed.~Very high energy interactions probe physics at ultra-short distance intervals, $\lambda \propto 1/E$.~Thus, to probe physics at small distance intervals, particularly the nature of space and time, we need to go to ultrahigh energies.~Cosmic $\gamma$-rays and neutrinos and cosmic-ray nucleons provide the highest observable energies in the universe.

This paper reviews the topic of testing the exactness of Lorentz symmetry using astrophysical observations of ultrahigh energy cosmic rays, $\gamma$-rays and neutrinos.~(For a review of tests of Lorentz symmetry in the gravitational sector, see~\cite{he16}.)~Cosmic $\gamma$-rays have been observed at energies up to~$\cal{O}$$(10^3)$ GeV. Neutrinos of astrophysical origin have been observed at energies up~to~$\cal{O}$$(10^6)$~GeV. Even ultrahigh energy cosmic ray nuclei have been observed only up to $\cal{O}$$(10^{11})$~GeV energy.~These~energies are much lower than $M_{Pl}$.~However, the effects of violations of Lorentz invariance
(LIV) can be manifested at energies much lower than $M_{Pl}$. (By Noether's theorem, Lorentz symmetry implies Lorentz invariance; here we use these terms interchangeably.)

There are many ways that LIV terms in the free particle Lagrangian can affect physics at astrophysically-accessible energies.~In particular, the effect of LIV can change the kinematics of particle interactions and decays and also the threshold energies for these processes~\cite{co99,sg01}. LIV can also modify neutrino oscillations~\cite{co99,st17}. 


\section{The Coleman--Glashow Formalism} 

As an initial framework for exploring the effects of LIV, we start with 
the phenomenological approach of Coleman and Glashow~\cite{co99}.~This formalism has the advantage of (1) simplicity, (2)~preserving the $SU(3) \otimes SU(2) \otimes U(1)$ standard model of 
strong and electroweak interactions, (3) having the perturbative 
term in the Lagrangian
consist of operators of mass dimension four that thus preserves power
counting renormalizability and (4) being rotationally invariant in a
preferred frame that can be taken to be the rest frame of the CBR (Cosmic background radiation).
Coleman and Glashow start with a standard-model free-particle Lagrangian,
\begin{equation}
{\cal L} = \partial_{\mu} \Psi ^ * {\bf Z} \partial^{\mu}\Psi - \Psi ^
* {\bf M}^2\Psi
\end{equation}
\noindent where the matrices are in boldface, $\Psi$ is a column vector of $n$ fields with U(1)
invariance, and the positive Hermitian matrices ${\bf Z}$, and {\bf
M}$^2$ can be transformed so that ${\bf Z}$ is the identity and {\bf
M}$^2$ is diagonalized to produce the standard theory of $n$ decoupled
free fields. (We adopt the standard notation of denoting four-vector
indexes by Greek letters and three-vector spatial indexes by Latin letters).

They then add a leading order perturbative, Lorentz violating term
constructed from only spatial derivatives with rotational symmetry so
that:
\begin{equation}
{{\cal L} \rightarrow {\cal L} + \partial_i\Psi \eta
\partial^i\Psi} ,
\end{equation}
\noindent where {\bf $\eta$} is a small dimensionless Hermitian matrix
that commutes with {\bf M}$^2$ so that the fields remain separable and
the resulting single particle energy-momentum eigenstates go from
eigenstates of {\bf M}$^2$ at low energy to eigenstates of {\bf
$\eta$} at high energies.

To leading order, this term shifts the poles of the propagator,
resulting in the free particle dispersion relation:
\begin{equation}
E^2 = \vec{p} \ ^{2} + m^{2} + \eta \vec{p} \ ^2.
\label{dispersion}
\end{equation}

 This can be put in the standard form for the dispersion relation:

\begin{equation}
E^2 = \vec{p \ }{^2}c_{MAV}^2 + m^{2} c_{MAV}^4,
\end{equation}

\noindent by shifting the renormalized mass by the small amount $m
\rightarrow m/(1+\eta)$ and shifting the velocity from $c \ (= 1)$ by
the amount $c_{MAV} = \sqrt {(1 + \eta)} \simeq 1 + \eta/2$. The group 
velocity is given by:
\begin{equation}
{{\partial E}\over{\partial |\vec{p}|}} = {{|\vec{p}|} \over {\sqrt
{|\vec{p}|^2 + m^2 c_{MAV} ^2}}} c_{MAV},
\label{MAV}
\end{equation}

\noindent which goes to $c_{MAV}$ in the limit of large $|\vec{p}|$.~Thus, Coleman and Glashow identify $c_{MAV}$ to be the maximum
attainable velocity of the free particle.~Using this formalism, it
becomes apparent that, in~principle, different particles can have
different maximum attainable velocities (MAVs), which can be different
from $c$. Hereafter, we denote the MAV of a particle of type $i$ by
$c_{i}$ and the difference:
\begin{equation}
c_{i} - c_{j}~=~ {{\eta_{i}-\eta_{j}}\over{2}}~ \equiv~
\delta_{ij}.
\label{deltadef}
\end{equation}

\section{LIV Modified Kinematics for QED} 

Using the formalism of~\cite{co99}, we denote
maximum attainable 
velocity (MAV) of a particle of type $i$ by $c_{i}$.~We further define 
the difference $c_{i} - c_{j}
\equiv \delta_{ij}$, and specifically, here, $c_{e\gamma} \equiv \delta \ll c$
where $c \equiv 1$ is the low energy photon velocity. 
These definitions will be used to discuss the physics implications of
cosmic ray and cosmic $\gamma$-ray observations~\cite{sg01}. 

If $\delta < 0$, 
the decay of a photon into an electron-positron pair is kinematically allowed
for photons with energies exceeding $E_{\rm max}= m_e\,\sqrt{2/|\delta|}$.~This decay would take place rapidly, so that photons with energies 
exceeding $E_{\rm max}$ could not be observed either in the laboratory or as 
cosmic rays. Since~photons have been observed with energies 
$E_{\gamma} \ge$ 50~TeV from the Crab Nebula \cite{ta98}, this implies 
that $E_{\rm max}\ge 50\;$TeV, or that $|\delta| < 2\times 
10^{-16}$.
 
If, on the other hand, $\delta > 0$, electrons become 
superluminal if their energies exceed $E_{\rm max}/\sqrt{2}$.
Electrons traveling faster than light will emit light at all frequencies by a
process of `vacuum \v{C}erenkov radiation'.~The
electrons then would rapidly lose energy until they become subluminal.~Because electrons have been seen in the cosmic radiation 
with energies up to $\sim\,$2~TeV~\cite{ni80}, it follows that 
$\delta < 3 \times 10^{-14}$.~A smaller, but more indirect, upper limit on $\delta$ for the $\delta > 0$
case can be obtained from theoretical considerations of $\gamma$-ray emission 
from the Crab Nebula. Its emission above 0.1 GeV, extending into the TeV range, 
is thought to be Compton emission of the same relativistic electrons that
produce its synchrotron radiation at lower energies~\cite{ah01}.~The Compton component, 
extends to 50 TeV and thus implies the existence of electrons having energies 
at least this great
in order to produce 50 TeV photons, even in the extreme Klein-Nishina limit.
This indirect argument, based on the reasonable assumption that the 50-TeV
$\gamma$-rays are from Compton interactions, leads to a smaller upper limit on 
$\delta$, viz., $\delta < 10^{-16}$.~A further constraint on $\delta$ for $\delta > 0$ 
follows from a change in the threshold energy for the pair 
production process $\gamma + \gamma \rightarrow e^+ + e^-$. 
This follows from the fact that the square of the 
four-momentum is changed to give the threshold condition:

\begin{equation}
2\epsilon E_{\gamma}(1-cos\theta)~ -~ 2E_{\gamma}^2\delta~\ge~ 4m_{e}^2,
\label{absthresh}
\end{equation}

\noindent where $\epsilon$ is the energy of the low energy photon and 
$\theta$ is the
angle between the two photons. The second term on the left-hand side comes
from the fact that $c_{\gamma} = \partial E_{\gamma}/\partial p_{\gamma}$.
It follows that the condition for a~significant increase in the energy
threshold for pair production is $E_{\gamma}\delta/2$ $ \ge$
$ m_{e}^2/E_{\gamma}$, or 
equivalently, 
\begin{equation}\delta \ge {2m_{e}^{2}/E_{\gamma}^{2}}.
\label{absdelta}
\end{equation}

The $\gamma$-ray spectrum of the active galaxy Mkn 501 while flaring 
extended to $E_{\gamma} \ge 24$ TeV \cite{ah01} and exhibited the 
high energy absorption expected from $\gamma$-ray annihilation by extragalactic pair-production interactions with extragalactic infrared 
photons~\cite{st92, ds02, ko03}.
This has led Stecker and Glashow \cite{sg01} to point out that the Mkn 501 
spectrum presents evidence for pair-production with no indication of 
Lorentz invariance violation (LIV) up to a photon energy of 
$\sim$20~TeV and to thereby place a quantitative constraint on LIV
given by $\delta < 2m_{e}^{2}/E_{\gamma}^{2} \simeq 
10^{-15}$.~This constraint on positive $\delta$ is 
more secure than the smaller, but indirect, limit given above. See~\cite{ja08a} 
for further considerations regarding the modification of \gray~ spectra by LIV.

\section{Ultrahigh Energy Cosmic Rays}
\subsection{Extragalactic Origin}

Ultrahigh energy cosmic rays (UHECRs) appear to have an isotropic distribution in arrival direction. Because of their ultrahigh
energy, their propagation is little affected by galactic magnetic
fields. Thus, owing to their observed isotropy, cosmic rays above 
10 EeV ($10^{19}$ eV) are believed to be of extragalactic origin. 

\subsection{The Greisen--Zatsepin--Kuz'min Effect}

Shortly after the discovery of the 2.7 K CBR, Greisen \cite{gr66} and Zatsepin and Kuz'min \cite{za66} predicted that pion-producing interactions of UHECR protons
with the microwave photons of the CBR should produce a spectral cutoff at $E \sim$ 50 EeV. The
flux of UHECRs is thus expected to be
attenuated by such meson-producing interactions. This effect is generally known as the ``GZK (Greisen--Zatsepin--Kuz'min) effect''. Owing to these interactions, extragalactic protons with energies above $\sim$100~EeV should be attenuated while propagating from distances
beyond $\sim$$100$ Mpc~\cite{st68}. The cross-section for this process
peaks at the $\Delta$ baryon resonance, 
\begin{equation}
p + \gamma \rightarrow \Delta \rightarrow N\pi .
\label{photopi} 
\end{equation}

The energy threshold
for photomeson interactions of UHECR protons of initial laboratory
energy $E$ with low energy photons of the 2.7 blackbody CBR having 
laboratory energy $\epsilon$ is determined by the relativistic invariance 
of the square of the total four-momentum of the proton-photon 
system. This~relation, together with the threshold inelasticity 
relation $E_{\pi} = m/(M + m) E$ for single pion production, 
yields the threshold conditions for head on collisions in the laboratory frame:
\begin{equation}
4\epsilon E = m(2M + m)
\end{equation}

\noindent for the proton, and:
\begin{equation}
4\epsilon E_{\pi} = {{m^2(2M + m)} \over {M + m}}
\label{pion}
\end{equation}
\noindent for the pion, where $M$ is the rest mass of the
proton and $m$ is the rest mass of the pion~\cite{st68}.

\section{Modification of the GZK Effect }
\label{GZKmod}
We have previously discussed the observations of UHECRs and their
expected attenuation over extragalactic distances owing to the
GZK effect. We now turn our attention to how the GZK effect can
be modified by a very small violation of Lorentz symmetry.

\subsection{Energy Threshold}

If Lorentz symmetry is slightly broken so that $c_\pi~ >~ c_p$, it follows from Equations (\ref{deltadef}), 
(\ref{pion}) and~(\ref{dispersion}) and that the
threshold energy for photomeson is altered because the square of the
four-momentum is shifted from its LI form so that the threshold
condition in terms of the pion energy becomes:
\begin{equation}
4\epsilon E_{\pi} = {{m^2(2M + m)} \over {M + m}} + 2 \delta_{\pi p}
E_{\pi}^2
\label{LIVpi}
\end{equation}

\noindent 

Equation (\ref{LIVpi}) is a quadratic equation with real roots only
under the condition:
\begin{equation}
\delta_{\pi p} \le {{2\epsilon^2(M + m)} \over {m^2(2M + m)}} \simeq
\epsilon^2/m^2.
\label{root}
\end{equation}

Defining $\epsilon_0 \equiv kT_{CBR} = 2.35 \times 10^{-4}$ eV with
$T_{CBR} = 2.725\pm 0.02$ K, Equation (\ref{root}) can be rewritten:
\begin{equation}
\delta_{\pi p} \le 3.23 \times 10^{-24} (\epsilon/\epsilon_0)^2.
\label{CG}
\end{equation}

Equation (\ref{CG}) implies that if Lorentz symmetry is violated and 
for $\delta_{\pi p} > 0$, photomeson interactions with CBR photons
will not occur for $\delta_{\pi p} \ge \sim 3 \times 10^{-24}$. 
This condition, together with Equation (\ref{LIVpi}), implies 
that while photomeson interactions leading to
GZK suppression can occur for ``lower energy'' UHE protons interacting
with higher energy CBR photons on the Wien tail of the CBR blackbody 
spectrum, other
interactions involving higher energy protons and photons with smaller
values of $\eta$ will be forbidden. Thus, the observed UHECR
spectrum may exhibit the characteristics of GZK suppression near the
normal GZK threshold, but the UHECR spectrum can ``recover'' at higher
energies owing to the possibility that photomeson interactions at
higher proton energies may be forbidden. 

\subsection{Detailed GZK Kinematics with LIV}

We have seen that kinematical relations governing photomeson interactions are
changed in the presence of even a small violation of Lorentz
invariance. We now consider this in more detail~\cite{ss09}.

Following Equations (\ref{deltadef}) and (\ref{dispersion}), we denote:

\begin{equation}
E^2=p^2+2\delta _a p^2 +{m_a}^2
\label{dispersiona}
\end{equation}
\noindent where $\delta _a$ is the difference between the MAV for the
particle {\it a} and the speed of light in the low momentum limit ($c
= 1$).

In the cms (center-of-momentum system), the
energy of particle $a$ is then given by:
\begin{equation}
\sqrt{s_a} = \sqrt{E^2-p^2} = \sqrt{2\delta_a p^2 + m_a^2} \ \ge \ 0.
\label{restmass}
\end{equation}

Owing to LIV, in the cms, the particle will not generally be at rest
when $p = 0$ because:
\begin{equation}
v = {{\partial E} \over {\partial p}} \neq {{E}\over {p}}.
\end{equation}

The modified kinematical relations containing LIV have a strong effect
on the amount of energy transferred from an incoming proton to the pion
produced in the subsequent interaction, i.e., the~inelasticity
\cite{ss09,al03}.~The total inelasticity, $K$, is an average of
$K_{\theta}$, which depends on the angle between the proton and photon
momenta, $\theta$:
\begin{equation}
K = \frac{1}{\pi }\int\limits_0^\pi {K_\theta d\theta }.
\label{Ktot}
\end{equation}

The primary effect of LIV on photopion production is a reduction of
phase space allowed for the interaction. This results from the limits
on the allowed range of interaction angles integrated over in order to
obtain the total inelasticity from Equation (\ref{Ktot}). For
real-root solutions for interactions involving higher energy protons,
the range of kinematically allowed angles in Equation (\ref{Ktot})
becomes severely restricted. The modified inelasticity that results
is the key in determining the effects of LIV on photopion
production. The inelasticity rapidly drops for higher incident proton
energies.

\begin{figure}
\centering
\includegraphics[width=9.2cm]{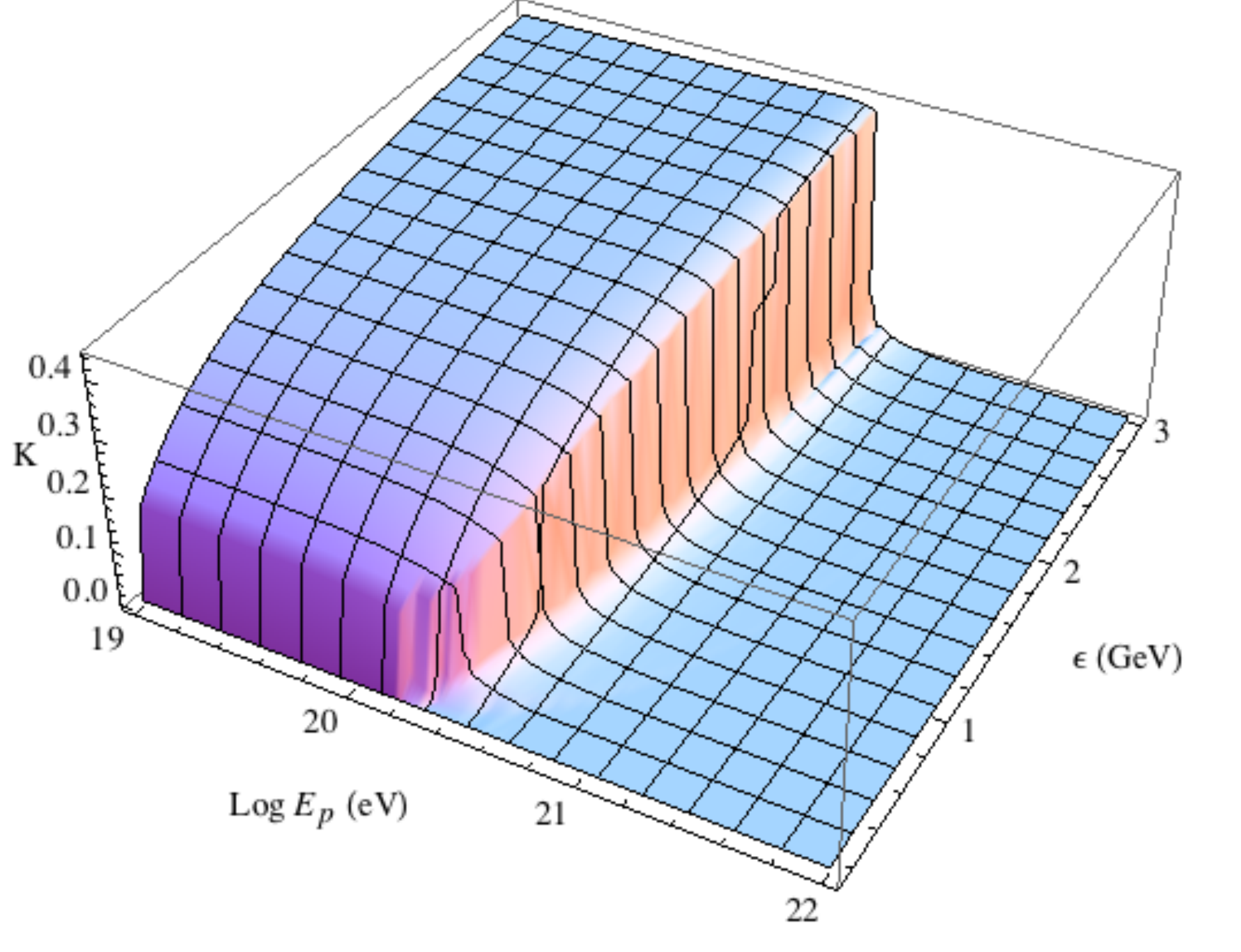}
\caption{The calculated proton inelasticity modified by LIV for
$\delta_{\pi p} = 3 \times 10^{-23}$ as a function of CBR photon
energy and proton energy \protect \cite{ss09}.}
\label{inelasticity}
\end{figure}

Figure \ref{inelasticity} shows the calculated proton inelasticity
modified by LIV for a value of $\delta_{\pi p} = 3 \times 10^{-23}$ as
a function of both CBR photon energy and proton energy \cite{ss09}.
Other choices for $\delta_{\pi p}$ yield similar plots. The principal
result of changing the value of $\delta_{\pi p}$ is to change the
energy at which LIV effects become significant.~For a choice of
$\delta_{\pi p} = 3 \times 10^{-23}$, there is no observable effect
from LIV for $E_{p}$ less than $\sim200$ EeV. Above this energy, the
inelasticity precipitously drops as the LIV term in the pion rest
energy approaches $m_{\pi}$.

With this modified inelasticity, the proton energy loss rate by
photomeson production is given by:

\begin{equation}
{{1}\over{E}}{{dE}\over{dt}}  =  -  {{\epsilon_{0}c}\over{2\pi^2
\gamma^2}\hbar^3c^3} \int\limits_{\epsilon_{th}}^\infty d\epsilon^{\prime}~ \epsilon^{\prime}~
\sigma(\epsilon^{\prime})                   K(\epsilon^{\prime})
\ln[1-e^{-\epsilon^{\prime}/2\gamma\epsilon_{0}^{\prime}}]
\label{modified}
\end{equation}

\noindent where $\epsilon_{th}$ is the photon threshold energy for the
interaction and $\sigma(\epsilon^{\prime})$ is the total $\gamma$-p
cross-section with contributions from direct pion production,
multipion production and the $\Delta$ resonance.

\begin{figure}
\centering
\includegraphics[width=8 cm]{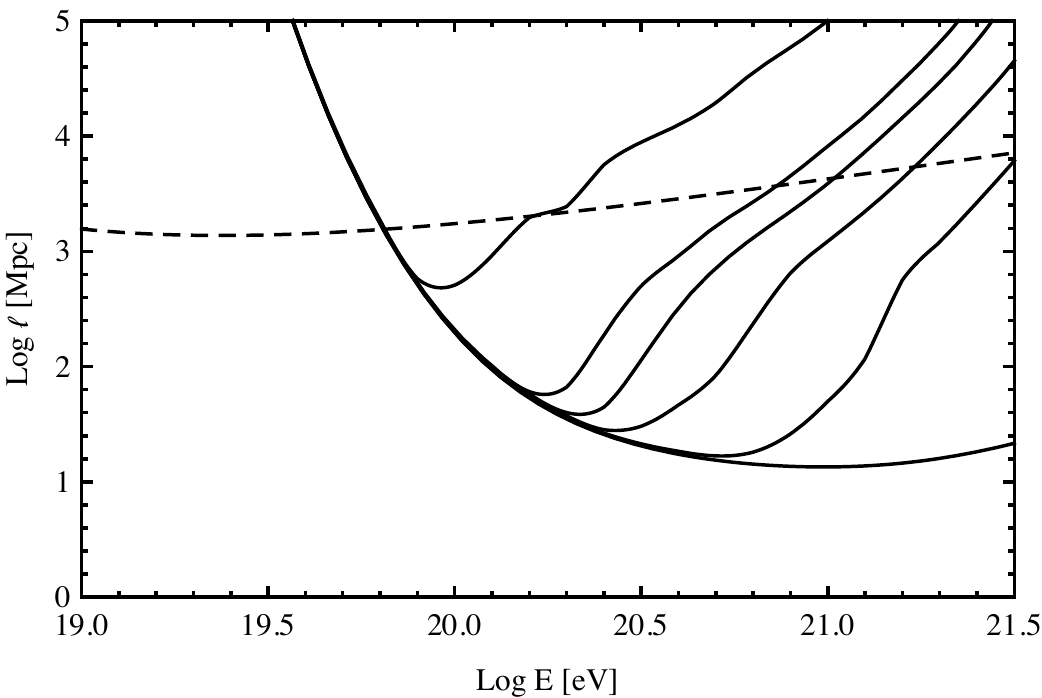}
\caption{The calculated proton attenuation lengths as a function of
proton energy modified by LIV for various values of $\delta_{\pi p}$
(solid lines), shown with the attenuation length for pair production
unmodified by LIV (dashed lines). From top to bottom, the curves are
for $\delta_{\pi p} = 1 \times 10^{-22}$, $3~\times~10^{-23}$, $2~\times~10^{-23}$, $1~\times~10^{-23}$, $3 \times 10^{-24}$, 0 (no Lorentz
violation) \protect \cite{ss09}.}
\label{attenuation}
\end{figure}

The corresponding proton attenuation length is given by $\ell =
cE/r(E)$, where the energy loss rate $r(E) \equiv (dE/dt)$. This
attenuation length is plotted in Figure \ref{attenuation} for various values
of $\delta_{\pi p}$ along with the unmodified pair production
attenuation length from pair production interactions, $p +
\gamma_{CBR} \rightarrow e^+ + e^-$.






\section{Comparison of LIV-Modified UHECR Spectra with Observations}

An analytic calculation of LIV-modified UHECR proton spectra was obtained using Equation~(\ref{modified})~\cite{ss09}~(see also \cite{al03}). In order to take account of the
probable redshift evolution of UHECR production in astronomical
sources, one must take account of the following considerations~\cite{ss09}: 
\\
\\	
1. The CBR photon number density increases as
$(1+z)^3$, and the CBR photon energies increase linearly with $(1+z)$.
The corresponding energy loss for protons at any redshift $z$ is thus
given by:
\begin{eqnarray}
r_{\gamma p}(E,z) = (1+z)^3 r[(1+z)E].
\label{eq5}
\end{eqnarray}

\noindent 2. It is assumed that the average UHECR volume
emissivity is of the form given by
\mbox{$q(E_i,z) = K(z)E_i^{-\Gamma}$} where $E_i$ is the initial energy of
the proton at the source and $\Gamma = 2.55$. 
The source evolution is assumed to be $K(z) \propto (1 + z)^{3.6}$ with $z \le 2.5$ so
that $K(z)$ is roughly proportional to the empirically-determined
$z$-dependence of the star formation rate. $K(z=0)$ and $\gamma$ are
normalized to fit the data below the GZK threshold.
\\

\begin{figure}
\centering
\includegraphics[width=9.5cm]{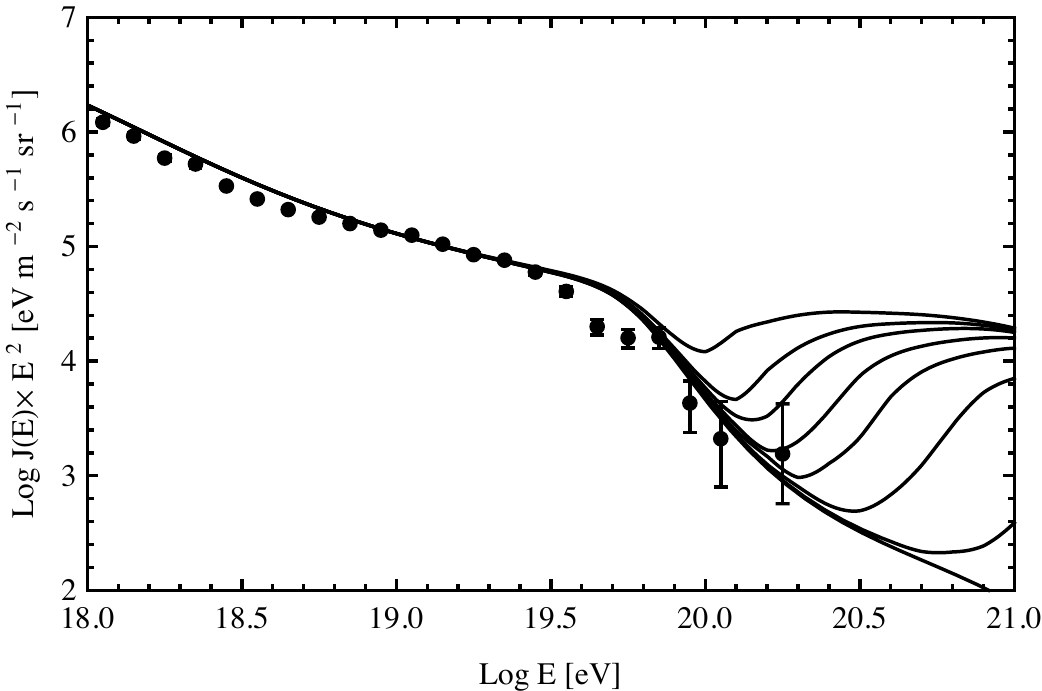}
\caption{Comparison of Auger data \protect \cite{sch09} with calculated spectra
for various values of $\delta_{\pi p}$, taking $\delta_p = 0$ (see the
text). From top to bottom, the curves give the predicted spectra for
$\delta_{\pi p} = 1 \times 10^{-22}$, $6 \times 10^{-23}$, $4.5 \times
10^{-23}, 3 \times 10^{-23}$, $2 \times 10^{-23}$, $1 \times 10^{-23}$, $3
\times 10^{-24}$, 0 (no Lorentz violation) \protect \cite{ss09}.}
\label{Auger}
\end{figure}

The curves
calculated in \cite{ss09} assuming various values of $\delta_{\pi
p}$ are shown in Figure \ref{Auger} along with the {Auger} data
from \cite{sch09}.~They show that {even a very small amount
of LIV that is consistent with both a GZK effect and with the present
UHECR data can lead to a ``recovery'' of the UHECR spectrum at higher~energies.}

\section{Modified GZK Neutrino Spectrum from a Modified UHECR Spectrum}

Subsequent to the photopion interactions of Equation (\ref{photopi}), the charged pions 
that are produced decay to produce neutrinos.
It follows from the modified kinematics and inelasticity for photopion interactions as presented in Section \ref{GZKmod} (see Figure \ref{inelasticity}) that there will be a modification of the spectrum of the ``GZK neutrinos'' that result. 

\begin{figure}
\begin{center}
\includegraphics[height=2.5in]{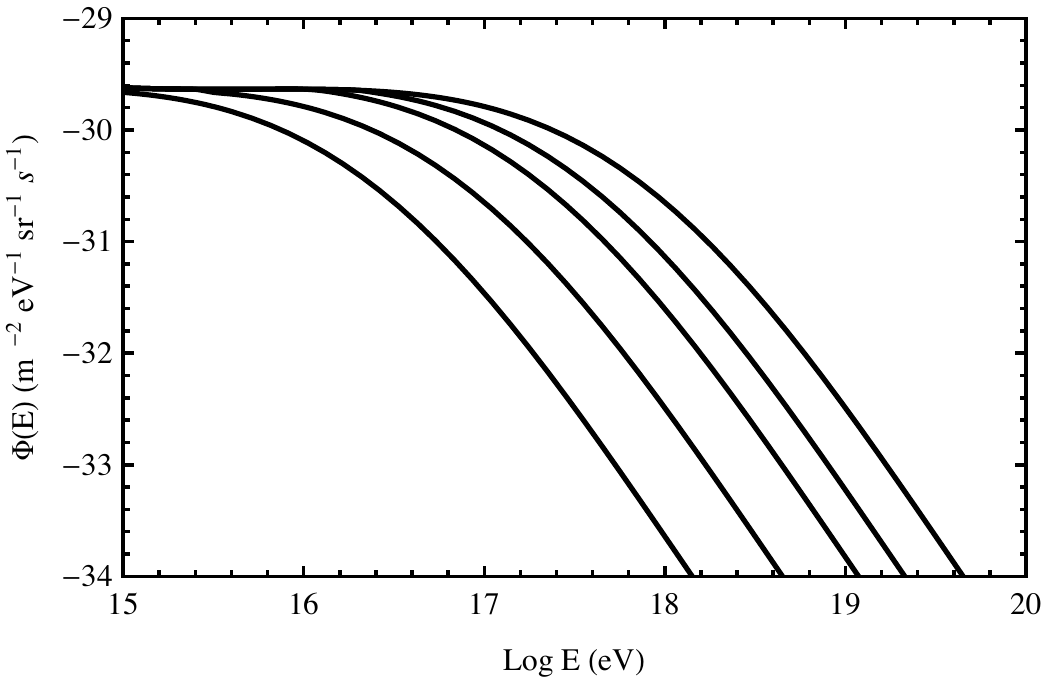}
\end{center}
\caption{Neutrino fluxes (of all species) corresponding to the UHECR models considered in Figure~\ref{Auger}. From left to right, the curves give the predicted fluxes for $\delta_{\pi p} = 1 \times 10^{-22}$, $6 \times 10^{-23}$, $3 \times 10^{-23} $, $1 \times 10^{-23}$, 0~\cite{ss11}.}
\label{neutrinoflux}
\end{figure}

Figure \ref{neutrinoflux} shows the corresponding total neutrino flux (all species) for the same choices of $\delta_{\pi p}$ as the UHECR spectra presented in figure \ref{Auger}.~As expected, increasing $\delta_{\pi p}$ leads to a decreased flux of higher energy photomeson neutrinos because the interactions involving higher energy UHECRs are suppressed \cite{ss11}. It is also evident that the peak energy of the neutrino energy flux spectrum (EFS), $E\Phi(E)$, shifts to lower energies with increasing $\delta_{\pi p}$.  
This possible LIV effect on GZK neutrinos may be tested using radio arrays built on Antarctic ice that are based on the Askaryan effect such as the Askaryan Radio Array (ARA)
~\cite{al16} and the Antarctic Ross Ice-Shelf Antenna Neutrino Array (ARIANNA)
~\cite{ba17}. GZK effect neutrinos can also be observed by space-based telescopes looking down at the atmosphere to study the horizontal air showers they produce~\cite{st04,ol17}.

\section{Generalizing the Coleman--Glashow Formalism}

The Coleman--Glashow formalism preserves a renormalizable Lagrangian structure, with the Lagrangian density being of mass-dimension $[d]
 = 4$. More generally, it is possible to construct an extended standard model (SME) Lagrangian as an effective field theory formalism that is testable and that holds at energies $E \ll M_{Pl}$~\cite{ko89}. This formalism admits LIV terms composed of operators of dimension $[d] > 4$ suppressed by values of $M_{Pl}^{(n-4)}$. It is motivated by two ideas: (1) any LIV extant at accelerator energies must be extremely small and (2) physics must be strongly modified at the Planck energy, $M_{Pl}$. These considerations are reflected by adding small Lorentz-violating terms in the free particle Lagrangian that are suppressed by powers of some quantum gravity energy scale $E_{QG}$ $\sim$ ${\cal{O}}(M_{Pl})$. 
Sums of such terms are also possible and, as such, admit energy-dependent values of $\delta$,
\begin{equation}
\delta = \sum_{n=0,1,2}\lambda_n\left(\frac{E}{M_{Pl}}\right)^n.
\label{d}
\end{equation}
\noindent where we will consider only terms up to \emph{n} = 2. In the SME formalism, the \emph{n} = 1 term corresponds to the lowest order $\cal{CPT}$-odd operator, and the \emph{n} = 2 term corresponds to the lowest order $\cal{CPT}$-even operator. We now consider two applications of this formalism.

\subsection{Time of Flight from \gray~Bursts}

Energy-dependent MAVs constructed by assuming that LIV is related to physics at the Planck scale have been used in time-of-flight observations of photon arrival times from cosmologically-distant \gray~ bursts. For this analysis, LIV modifications to the photon dispersion relation were taken to be the sum of a series of terms in $E/E_{QG}$~\cite{ja08},
\begin{equation}
\label{dispersion}
\ensuremath{E^{2}\simeq p^{2}\times\left[1 -\overset{n=2}{\underset{n=1}{\sum}}\mathit{s_{\pm}}\left(\frac{E}{E_{QG}}\right)^{n}\right]},
\end{equation}
\noindent where it was expected that $E_{QG}$ = ${\cal{O}}$$(M_{Pl})$. In Equation (\ref{dispersion}), $\mathit{s_\pm}$ is the ``sign of LIV'', taken equal to $+$1 ($-$1) for a decrease (increase) in photon MAV with an increasing photon energy. In case the $n=1$ term is suppressed, owing to conservation of $\cal{CPT}$, the $n = 2$ is assumed to dominate.

Adopting the notation of~\cite{vv13}, we define an``LIV parameter'' $\tau_n$ as the ratio of the time delay between photons of the highest observed energy, $E_h$, and the lowest observed energy,~$E_l$, and the quantity $(E_h^n$ $-$ $E_l^n)$. Then:
\begin{equation}
\label{deltatn}
\tau_n 	\equiv \frac{\Delta t}{E_{h}^n - E_{l}^n} \simeq \mathit{s_\pm} \frac{(1+n)}{2H_0}\frac{1}{E_{QG}^n}\times \kappa_n,
\end{equation}
\noindent where:
\begin{equation}
\label{eq:kappa}
 \kappa_n\equiv\int\limits_0^z\frac{(1+z')^n}{\sqrt{\Omega_{\Lambda} + \Omega_{\rm M} (1 + z')^3}}dz'
\end{equation}
\noindent where $\kappa_n$ is a cosmological distance parameter. (Values of $\Omega$ are conventionally defined as fractions of the critical density needed to close the universe, $3H_{0}/8\pi G$.) Here, as usual, $z$ is redshift; $H_{0}$ is the Hubble constant; and values of $\Omega_{\lambda}$ and $\Omega_m$ are designated for the dark energy density and total matter density, respectively.~The Hubble constant is taken to be $H_0 =$ 67.8 km s$^{-1}$ Mpc$^{-1}$; \mbox{$\Omega_{\Lambda}$ = 0.7}; and~$\Omega_{\rm M}$ = 0.3. 

Three different statistical analyses on Fermi observations of three $\gamma$-ray bursts have been used to derive strong constraints on $\tau_n$~\cite{vv13}.~The most stringent limits for $s = +1$~at the 95\% confidence level
are obtained from the $\gamma$-ray burst GRB~090510 and are $E_{QG} > 7.6 M_{Pl}$ and $E_{QG} > 1.3 \times 10^{11}$~GeV for linear and quadratic leading order LIV-induced vacuum dispersion, respectively. 

An early suggestion for testing quantum gravity~\cite{ac98} employed the natural idea of identifying $E_{QG} \equiv M_{Pl}$. A later ``D-brane'' model, inspired by string theory concepts, suggested the prospect that only photons would exhibit an energy-dependent velocity~\cite{el08}. That model envisions a universe filled with a gas of point-like D-branes that only interact with photons.~It predicts that vacuum has an energy-dependent index of refraction that causes only a retardation, again with $E_{QG} \simeq M_{Pl}$. Such~models are disfavored by the analysis of the { Fermi} observations that, on its face, appears to suggest $E_{QG} > 7.6 M_{Pl}$.

It is, of course, unphysical to invoke the interpretation $E_{QG} > M_{Pl}$, as implied in the $n = 1$ case. However, if we write the first-order dispersion relation for the group velocity of a photon with energy~$E$~as: 
\begin{equation}
\label{eq:uph}
c_{\gamma}(E)=\frac{\partial E}{\partial p}\simeq \left[1-\lambda_1 {{E}\over{M_{Pl}}}\right],
\end{equation}
in analogy with Equations (\ref{MAV}) and (\ref{d}), we find the limit $\lambda_1 < 0.13$, in contrast with the prediction $\lambda_1 \simeq 1$. 

\subsection{Gamma-Ray Absorption with Planck Suppressed LIV}
\label{abs-n}

Motivated by the time-of-flight result above, we drop the condition from Equation (\ref{dispersion}) that $\lambda = \pm 1$ and consider $\lambda_n \neq 1$.~Using Equation (\ref{d}), the energy threshold for relation (\ref{absdelta}) is modified slightly in the Planck-suppressed LIV case.~If we again take the condition for LIV to significantly change the threshold energy for absorption as given by the relation (\ref{absdelta}), with $\delta$ given by Equation (\ref{d}), the criterion for an absorption feature in the \gray~ energy spectrum of a source to be affected by an~LIV term with Planck suppression becomes:
\begin{equation}
E_{\gamma} > \left({{2m_e^2M_{Pl}^n}\over{{\lambda_n}}}\right)^{1\over{n+2}}
\end{equation}
where $n$ is the lowest value for the LIV term that dominates in the series (\ref{d}). The $n = 1$ case was first treated in~\cite{ki99}. This topic has been considered further in~\cite{ja08a}. 

\section{Vacuum Birefringence}

Important fundamental constraints on LIV come from searches for the vacuum birefringence effect predicted within the framework of the effective field theory (EFT) analysis of~\cite{mp03}~(see also~\cite{ck98,km13})\footnote{Vacuum birefringence from LIV differs from the identically-labeled effect that can occur in a strong magnetic field from interactions with virtual electrons, causing the index of refraction in vacuo to be different for the two polarization modes, depending on their orientation to the magnetic field.}. Within this framework, applying the Bianchi identities to the leading order Maxwell equations in vacuo, a $\cal{CPT}$-odd mass dimension five operator term is derived of the form:
\begin{equation}
{{\cal{L}}_{\gamma} = {{\xi}\over{M_{Pl}}}{n^aF_{ad}n\cdot \partial (n_b {F}^{bd})}}
\label{bi}
\end{equation}

It is shown in~\cite{mp03} that the expression given in Equation (\ref{bi}) is the only dimension five modification of the free photon Lagrangian that preserves both rotational symmetry and gauge invariance.~This leads to a modification in the dispersion relation
proportional to \mbox{$\xi (\omega/M_{Pl})=\xi (E/M_{Pl})$}, with the new dispersion relation given by: 
\begin{equation}
\omega^2~=~ k^2 \pm \xi\, k^3/M_{Pl}. 
\label{disp-ph}
\end{equation}
\noindent with photons of opposite helicity having slightly different velocities. 
This leads to a rotation in the angle of polarization,
\begin{equation} 
\theta(t)=\left[\omega_+(k)-\omega_-(k)\right]t_{P}/2~\simeq~\xi k^2 t_{P}/2M_{Pl}
\label{rotation}
\end{equation}
for a plane wave with wave-vector $k$, where $\xi k/M_{Pl} \ll 1$ and where
$t_{P}$ is the propagation time. 

Observations of
polarized radiation from distant sources can thus be
used to place an upper bound on $\xi$. 
The vacuum birefringence constraint arises from the fact that
if the angle of polarization rotation (\ref{rotation})
were to differ by more than $\pi/2$ over the energy range
covered by the observation,
the instantaneous polarization
at the detector would fluctuate sufficiently
for the net
polarization of the signal to be suppressed well
below any observed value. 
The difference in rotation angles for wave-vectors $k_1$
and $k_2$ is:
\begin{equation}
 \Delta\theta=\xi (k_2^2-k_1^2) L_{P}/2M_{Pl},
 \label{diffrotation}
\end{equation}
where we have replaced the propagation time $t_{P}$ by the propagation distance $L_{P}$
from the source to the~detector. 

If polarization is detected from a source at redshift $z$, this yields the constraint: 
\begin{equation}
 |\xi|< {{\pi M_{Pl}}\over{\int\displaylimits_{0}^{z} dz'{[k_2(z')^2-k_1(z')^2]|dL_{P}(z')/dz'|}}}
\label{constraint}
\end{equation}
where $k_{1,2}(z') = (1+z')\cdot k_{1,2}(z'=0)$
and:
\begin{equation}
\Bigl{|}{dL_{P}\over{dz'}}\Bigr{|} = {{c}\over{H_{0}}}{{1}\over{(1+z')\sqrt{\Omega_\Lambda + (1+z')^3\Omega_m}}}.
\end{equation}

Present observations of hard X-rays from Mpc-distant \gray~bursts put limits on $\xi$ of order $10^{-15}$--$10^{-16}$, e.g.,~\cite{st11,go14}.

\section{LIV in the Neutrino Sector}

There are many ways that LIV terms in the free particle Lagrangian can affect neutrino physics. We~will here consider only one of these consequences, viz., the effect of resulting changes in the kinematics of particle interactions. (A discussion of the effect of LIV on modifying neutrino oscillations may be found in~\cite{st17}.)~These changes can modify the threshold energies for particle interactions, allowing or forbidding such interactions~\cite{co99,sg01}. 
Observational evidence for this effect can be searched for by examining the energy spectrum of high energy cosmic neutrinos obtained by the {IceCube} collaboration. We base our discussion of this evidence on the well delineated framework of the standard model extension (SME) formalism of effective field theory (EFT)~\cite{ck98}, as given in more detail in~\cite{st15}.

\subsection{Fermion LIV Operators with $[d] > 4$ LIV with Rotational Symmetry in SME}

We again assume rotational invariance in the rest frame of the CBR and consider only the effects of Lorentz violation on freely-propagating cosmic neutrinos. Thus, we only need to examine Lorentz-violating modifications to the neutrino kinetic terms. Majorana neutrino couplings are ruled out in SME in the case of rotational symmetry~\cite{ko12}.~Therefore, we only consider Dirac neutrinos. 

Using Equation (\ref{dispersion}), one can define an effective mass, $\tilde{m}(E)$, that is a useful parameter for analyzing LIV-modified kinematics. The effective mass is constructed to include the effect of the LIV terms. We define an effective mass $\tilde{m}_I(E)$ for a particle for type $I$ using the dispersion relation (\ref{dispersion}) as~\cite{co99, st15}: 
\begin{equation} 
\tilde{m}_{I}^2(E)=m_{I}^2+ 2\delta_I E_{I}^2,
\label{effectivemass}
\end{equation}
where the velocity parameters $\delta_I$ are now energy-dependent dimensionless coefficients for each species, $I$, that are contained in the Lagrangian.~Furthermore, we define the parameter $\delta_{IJ} \equiv \delta_{I} - \delta_{J}$ as the Lorentz-violating difference between the MAVs of particles $I$ and $J$. In general, $\delta_{IJ}$ will therefore be of the form given by Equation (\ref{d}).

If we wish to assume the dominance of Planck-suppressed terms in the Lagrangian as tracers of Planck-scale physics, it follows that that $\lambda_{\nu e,0} \ll \lambda_{\nu e,1}, \lambda_{\nu e,2}$. (Several mechanisms have been proposed for the suppression of the LIV $[d] = 4$ term in the Lagrangian. See, e.g., the review in Reference~\cite{li13}.) 
~Alternatively, we may {postulate} the existence of only Planck-suppressed terms in the Lagrangian, i.e., $\lambda_{\nu e,0} = 0$. We can further simplify by noting the important connection between LIV and $\cal{CPT}$ violation. Whereas a local interacting theory that violates $\cal{CPT}$ invariance will also violate Lorentz invariance~\cite{gr02}, the converse does not follow; an~interacting theory that violates Lorentz invariance may, or may not, violate $\cal{CPT}$ invariance. LIV~terms of odd mass dimension $[d]= 4 + n$ are $\cal{CPT}$-odd and violate $\cal{CPT}$, whereas terms of even mass dimension are $\cal{CPT}$-even and do not violate $\cal{CPT}$~\cite{ko09}. 

We can then specify a dominant term for $\delta_{IJ}$ in Equation (\ref{d}) depending on our choice of $\cal{CPT}$.~Considering Planck-mass suppression, the~dominant term that admits $\cal{CPT}$ violation is the $n = 1$ term in Equation (\ref{d}). On the other hand, if~we require $\cal{CPT}$ conservation, the
$n = 2$ term in Equation (\ref{d}) is the dominant term.~Thus,~we~can choose as a good approximation to Equation~(\ref{d}), a single dominant term with one particular power of $n$ by specifying whether we are considering $\cal{CPT}$-even or odd LIV. As a result, $\delta_{IJ}$ reduces to Equation~(\ref{d}) with $n = 1$ or $n = 2$ depending on the status of $\cal{CPT}$. We note that in the SME formalism, since $[d]$-odd LIV operators are $\cal{CPT}$-odd, the $\cal{CPT}$-conjugation property implies that neutrinos can be superluminal, while antineutrinos are subluminal or vice versa~\cite{ko12}. This will have consequences in interpreting our results, as we will discuss later.

In Equations (\ref{d}) and (\ref{dispersion}), we have not designated a helicity index on the $\lambda$ coefficients. The~fundamental parameters in the Lagrangian are generally helicity dependent.~In the $n = 1$ case, a~helicity dependence must be generated in the electron sector due to the $\cal{CPT}$-odd nature of the LIV~term. However, the constraints on the electron coefficient are extremely tight from observations of the Crab Nebula~\cite{st14b}. Thus, the contribution to $\lambda_{\nu e,1}$ from the electron sector can be neglected. In the $n = 2$ case, which is $\cal{CPT}$-even, we can set the left- and right-handed electron coefficients to be equal by imposing parity symmetry~\cite{st15}.

\subsection{LIV in the Neutrino Sector I: Lepton Pair Emission}
\label{vpe}

Since the Lorentz violating operators change the free field behavior and dispersion relation, interactions such as fermion-antifermion pair emission by slightly superluminal neutrinos become kinematically allowed~\cite{co99,co11} and can thus cause significant observational effects. An example of such an interaction is $\nu_e$ ``splitting'', i.e., $\nu_e \rightarrow \nu_e + \nu_i + \bar{\nu_i}$ where $i$ is a flavor index.~Neutrino splitting can be represented as a rotation of the Feynman diagram for neutrino-neutrino scattering, which is allowed by relativity. However, absent a violation of Lorentz invariance, neutrino splitting is forbidden by conservation of energy and momentum. In the the case of LIV-allowed superluminal neutrinos, the dominant pair emission reactions are neutrino splitting and its close cousin, vacuum electron-positron pair emission (VPE) $\nu_i \rightarrow \nu_i + e^+ + e^-$, as these are the reactions with the lightest final state masses. We now set up a simplified formalism to calculate the possible observational effect of these two~specific anomalous interactions on the interpretation of the neutrino spectrum observed by the~{IceCube}
collaboration.

\subsubsection{Lepton Pair Emission in the [\emph{d}] = 4 Case}

\begin{figure}
\begin{center}
\includegraphics[width=4.2in]{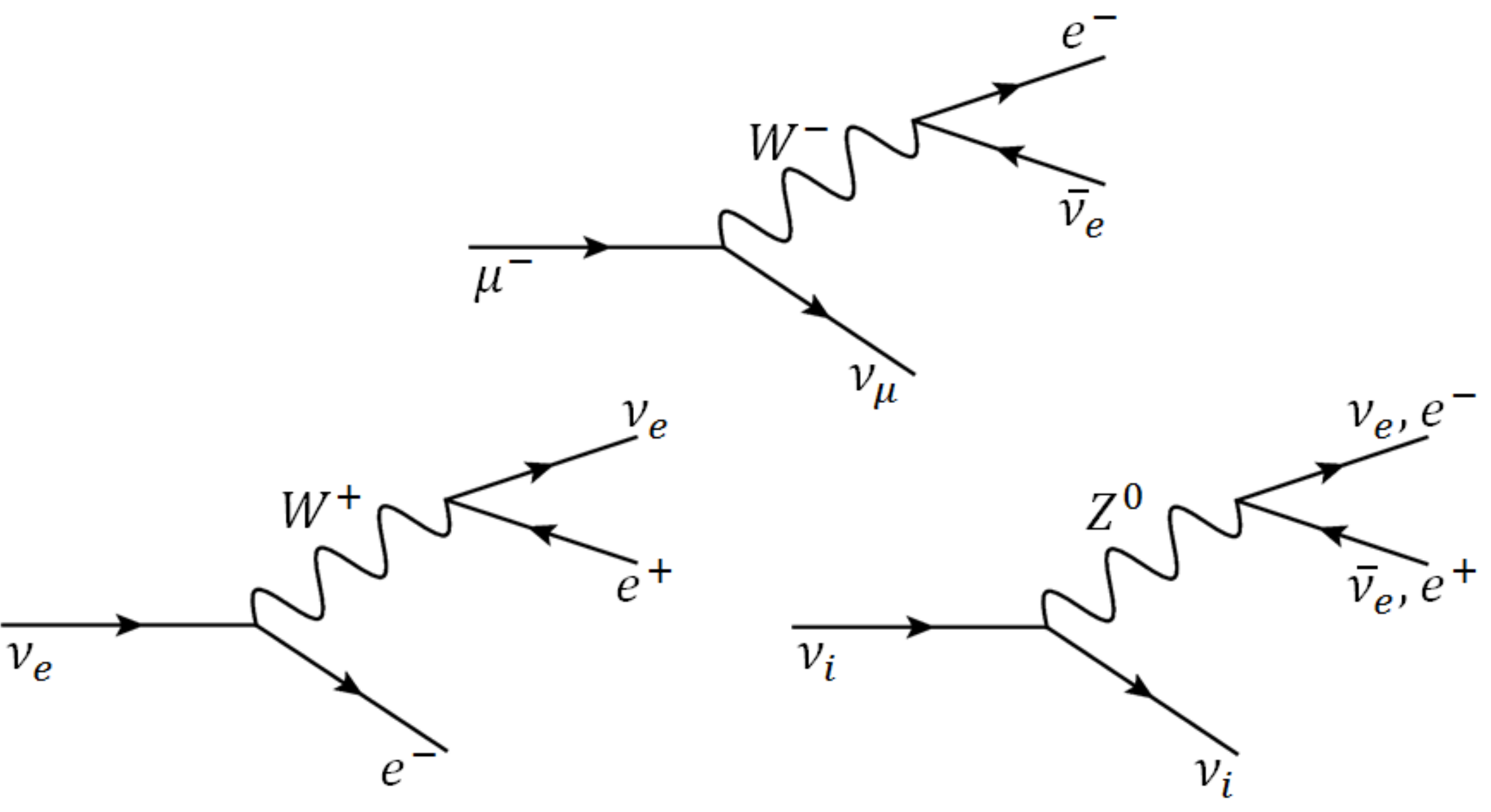}
\caption{Diagrams for muon decay (top), charged current-mediated vacuum electron-positron pair emission (VPE) (bottom left) and neutral current-mediated neutrino splitting and VPE (bottom right). Time runs from left to right, and the flavor index $i$ represents $e,\mu$, or $\tau$ neutrinos.}\label{fig:diagrams}
\end{center}
\end{figure}

In this section, we consider the constraints on the LIV parameter $\delta_{\nu e}$. We first relate the rates for superluminal neutrinos with that of a more familiar tree level, weak force-mediated standard model decay process: muon decay, $\mu^-\rightarrow \nu_\mu + \bar{\nu}_e + e^-$, as the processes are very similar (see Figure~\ref{fig:diagrams}). 

For muons with a Lorentz factor $\gamma_{\mu}$ in the observer's frame, the decay rate is found to be:
\begin{equation}
{ \Gamma \ = \gamma_{\mu}^{-1}} {{G_F^2 m_{\mu}^5}\over{192\pi^3}}
\label{mu}
\end{equation}
where $G_F^2 = g^4/(32M_{W}^4)$ is the square of the Fermi constant equal to $1.360 \times 10^{-10} \ {\rm GeV}^{-4}$, with $g$ being the weak coupling constant and $M_{W}$ being the $W$-boson mass in electroweak theory.

We apply the effective energy-dependent mass-squared formalism given by Equation (\ref{effectivemass}) to determine the scaling of the emission rate with the $\delta$ parameter and with energy.~Noting that for any reasonable neutrino mass, $m_{\nu} \ll 2\delta_{\nu e} E_{\nu}^2$, it follows that $\tilde{m_{\nu}}^2(E) \simeq 2\delta_{\nu e} E_{\nu}^2$.~(At relativistic energies, assuming that the Lorentz violating terms yield small corrections to $E$ and $p$, it follows that $E \simeq p$.)
 
We therefore make the substitution:
\begin{equation}
m_{\mu}^2 \ \rightarrow \ \tilde{m_{\nu}}^2(E) \simeq 2\delta_{\nu e} E_{\nu}^2 
\end{equation}
\noindent from which it follows that:
\begin{equation}
{\gamma_{\mu}^2} \ \rightarrow \ {{E_{\nu}^2}\over{2\delta_{\nu e} E_{\nu}^2}} = (2\delta_{\nu e})^{-1}.
\end{equation}

The rate for the vacuum pair emission processes (VPE) is then:
\begin{equation}
\Gamma \ \propto \ (2\delta_{\nu e})^{1/2} G_F^2 (2\delta_{\nu e} E_{\nu}^2)^{5/2} 
\label{dimension}
\end{equation}
\noindent which gives the proportionality:
\begin{equation}
\Gamma \ \propto \ G_F^2 \ \delta_{\nu e}^3 E_{\nu}^5 
\label{prop}
\end{equation}
\noindent showing the strong dependence of the decay rate on both $\delta_{\nu e}$ and $E_{\nu}$.

The energy threshold for $e^+e^-$ pair production is given by~\cite{sg01}:
\begin{equation}
E_{th} = m_e\sqrt{{{2}\over{{\delta_{\nu e}}}}} 
\label{threshold}
\end{equation}
with $\delta \equiv \delta_{\nu e}$ given by Equation (\ref{d}), the rate for the VPE process, $\nu \to \nu \,e^+\, e^-$, via the neutral current $Z$-exchange channel, has been calculated to be~\cite{co11}: 
\begin{equation}
\Gamma = \frac{1}{14}\frac{G_F^2 (2\delta)^3E_{\nu}^5}{192\,\pi^3} = 1.31 \times 10^{-14} \delta^3 E_{\rm GeV}^5\ \ {\rm GeV}.
\label{G}
\end{equation}
\noindent with the mean fractional energy loss per interaction from VPE of 78\%~\cite{co11}. 

In general, the charged current (CC) $W$-exchange channels contribute as well. However, this channel is only kinematically relevant for $\nu_e$'s, as the production of $\mu$ or $\tau$ leptons by $\nu_{\mu}$'s or~$\nu_{\tau}$'s has a much higher energy threshold due to the larger final state particle masses (Equation (\ref{threshold}) with $m_e$ replaced by $m_{\mu}$ or~$m_{\tau}$), with the neutrino energy loss from VPE being highly threshold dependent. Owing~to neutrino oscillations, neutrinos propagating over large distances spend 1/3 of their time in each flavor state. Thus, the flavor population of neutrinos from astrophysical sources is expected to be \mbox{[$\nu_e$:$\nu_{\mu}$:$\nu_{\tau}$] = [1:1:1]} so that CC interactions involving $\nu_e$'s will only be important 1/3 of the time. 

The vacuum \v{C}erenkov emission (VCE) process, $\nu \rightarrow \nu + \gamma$, is also kinematically allowed for superluminal neutrinos. However, since the neutrino has no charge, this process entails the neutral current channel production of a loop consisting of a virtual electron-positron pair followed by its annihilation into a photon. Thus, the rate for VCE is a factor of $\alpha$ lower than that for VPE.

\subsubsection{Vacuum $e^+e^-$ Pair Emission in the $[d] > 4$ Cases}

Using Equations (\ref{d}) and (\ref{prop}) and the dynamical matrix element taken from the simplest case~\cite{ca12}, we can generalize Equation (\ref{G}) for arbitrary values of $n=([d]- 4)$~\cite{st15}.
\begin{equation}
\Gamma = \frac{G_F^2}{192\,\pi^3}[(1-2s_W^2)^2 + (2s_W^2)^2]\zeta_n \lambda_n^3 \frac{E_\nu^{3n+5}}{M_{Pl}^{3n}} 
\label{G2}
\end{equation} 
\noindent where $s_W$ is the sine of the Weinberg angle ($s_W^2 = 0.231$) and the $\zeta_n$'s are numbers of order $1$~\cite{ca12}.
 
For the $n = 1$ case, we obtain the VPE rate:
\begin{equation}
\Gamma = 1.72 \times 10^{-14} \lambda_1^3 E_{\rm GeV}^5\ (E/M_{Pl})^3 \ {\rm GeV},
\label{Gn1}
\end{equation}
and for the $n = 2$ case, we obtain the VPE rate:
\begin{equation}
\Gamma = 1.91 \times 10^{-14} \lambda_2^3 E_{\rm GeV}^5\ (E/M_{Pl})^6 \ {\rm GeV}.
\label{Gn2}
\end{equation}
 
\section{LIV in the Neutrino Sector II: Neutrino Splitting}
\label{3n} 

The process of neutrino splitting in the case of superluminal neutrinos, i.e., $\nu \rightarrow 3\nu$, is relatively unimportant in the $[d] = 4, n = 0$ case~\cite{co11} owing to the small
velocity difference between neutrino flavors obtained from neutrino oscillation data~\cite{gg04, ab15,gg16}.~However, this is not true in the cases with \mbox{$[d] > 4$}.~In the presence of $[d] > 4$ $(n > 0)$ terms in a Planck-mass suppressed EFT, the velocity differences between the neutrinos, being energy dependent, become significant~\cite{mac13}. The daughter neutrinos travel with a smaller velocity.~The velocity-dependent energy of the parent neutrino is therefore greater than that of the daughter neutrinos.~Thus, the neutrino splitting becomes kinematically allowed. Let us then consider the $n = 1$ and $n = 2$ scenarios. 

Neutrino splitting is a neutral current (NC) interaction that can occur for all three neutrino flavors. The total neutrino splitting
rate obtained is therefore three times that of the NC-mediated VPE process above threshold.~Assuming the three daughter neutrinos
each carry off approximately 1/3 of the energy of the incoming neutrino, then 
for the $n = 1$ case, one obtains the neutrino splitting rate~\cite{st15}:
\begin{equation}
\Gamma = 5.16 \times 10^{-14} \lambda_1^3 E_{\rm GeV}^5\ (E/M_{Pl})^3 \ {\rm GeV},
\label{Gsplitn1}
\end{equation}
and for the $n = 2$ case, we obtain the neutrino splitting rate:
\begin{equation}
\Gamma = 5.73 \times 10^{-14} \lambda_2^3 E_{\rm GeV}^5\ (E/M_{Pl})^6 \ {\rm GeV}.
\label{Gsplitn2}
\end{equation}

The threshold energy for neutrino splitting is proportional to the neutrino mass so that it is negligible compared to that given by Equation ({\ref{threshold}).

\section{The Neutrinos Observed by {{IceCube}}} 
\label{ice}

As of this writing, the {IceCube} collaboration has identified $87_{-10}^{+14}$ events from neutrinos of astrophysical origin with energies above 10 TeV, with the error in the number of astrophysical events determined by the modeled subtraction of both conventional and prompt atmospheric neutrinos and also penetrating atmospheric muons at energies below 60 TeV~\cite{aa15}. 

There are are four indications that the bulk of cosmic neutrinos observed by {IceCube} with energies above 0.1 PeV are of extragalactic origin: (1) the arrival distribution of the reported events with $E > 0.1$ PeV observed by {IceCube} above atmospheric background is consistent with isotropy~\cite{aa13,aa14,aa15};
(2) at least one of the~PeV neutrinos came from a direction off the galactic plane~\cite{aa15};
(3) the diffuse galactic neutrino flux is expected to be well below that observed by {IceCube}~\cite{st79}; (4) upper limits on diffuse galactic $\gamma$-rays in the TeV--PeV energy range imply that galactic neutrinos cannot account for the neutrino flux observed by {IceCube}~\cite{ah14}.

Above 60 TeV, the {IceCube} data are roughly consistent with a spectrum given by $E_{\nu}^2(dN_{\nu}/dE_{\nu}) \simeq \ 10^{-8} \ {\rm GeV}{\rm cm}^{-2}{\rm s}^{-1}$~\cite{aa13,aa14,aa15}.~However, {IceCube} has not detected any neutrino-induced events from the Glashow resonance effect at $E_{\bar{\nu}_{e}} = M_W^2/2m_{e} = 6.3$ PeV~\cite{gl60}.~At the Glashow resonance energy electrons in the {IceCube} volume provide enhanced target cross-sections for $\bar{\nu}_{e}$'s through the $W^-$ resonance channel, $\bar{\nu}_{e} + e^- \rightarrow W^- \rightarrow shower$. This enhancement effect is expected to be about a factor of $\sim$$10$~\cite{aa13}. Owing to oscillations, it is expected that 1/3 of the potential 6.3 PeV neutrinos would be ${\nu}_{e}$'s plus $\bar{\nu}_{e}$'s unless new physics is involved.

Thus, the enhancement in the overall effective area expected is a factor of $\sim$3. Taking account of the increased effective area between 2 and 6 PeV and a decrease from an assumed neutrino energy spectrum of $E_{\nu}^{-2}$, we would expect about three events at the Glashow resonance energy, provided that the number of $\bar{\nu}_{e}$'s is equal to the number of ${\nu}_{e}$'s. Even without considering the resonance effect, several neutrino events above 2 PeV would be expected if the $E_{\nu}^{-2}$ spectrum extended to higher energies. Thus, the lack of this flux of neutrinos above $\sim$$2$ PeV energy and at the 6.3-PeV resonance may be indications of a cutoff in the neutrino spectrum\footnote{Since~\cite{st15} was published, an event with a~deposited energy of 2.6 PeV was reported by the IceCube collaboration. The energy of the neutrino that produced this event may have been as high as 6 PeV, although this is uncertain~\cite{aa16b}.}.

\section{Extragalactic Superluminal Neutrino Propagation}
\label{neutprop}
\label{cos}

Monte Carlo techniques have been employed to determine the effect of neutrino splitting and VPE on putative superluminal neutrinos~\cite{st15,st14b}.~Extragalactic neutrinos were propagated from cosmological distances taking account of the resulting energy loss effects by VPE and redshifting in the $[d] = 4$~case. In the $[d] > 4$ cases, energy loses include those from both neutrino splitting and VPE, as well as redshifting. It was assumed that the neutrino sources have a redshift distribution similar to that of the star formation rate~\cite{be13}. Such a redshift distribution appears to be roughly applicable for both active galactic nuclei and $\gamma$-ray bursts. A simple neutrino source spectrum proportional to $\sim$$E^{-2}$ was assumed between 100 TeV and 100 PeV, as is the case for cosmic neutrinos observed by {IceCube} with energies above 60 TeV~\cite{aa14}. The final results on the propagated spectrum were normalized to an energy flux of $E_{\nu}^2(dN_{\nu}/dE_{\nu}) \simeq 10^{-8}~{\rm GeV}~{\rm cm}^{-2}~{\rm s}^{-1}~{\rm sr}^{-1}$, consistent with the {IceCube} data for both the Southern and Northern Hemisphere. Results were obtained for VPE threshold energies between 10 PeV and 40~PeV as given by Equation ({\ref{threshold}), corresponding to values of $\delta_{\nu e}$ between $5.2\times10^{-21} \ {\rm and}~ 3.3 \times 10^{-22}$. 

Since the neutrinos are extragalactic and survive propagation from all redshifts, cosmological effects must be taken into account in deriving new LIV constraints. Most of the cosmic PeV neutrinos will come from sources at redshifts between $\sim$0.5 and $\sim$2~\cite{be13}. The effect of the cosmological $\Lambda$CDM redshift-distance relation is given by:
\begin{equation}
D(z) = {{c}\over{H_0}}\int\limits_0^z\frac{dz'}{(1+z')\sqrt{\Omega_{\Lambda} + \Omega_{\rm M} (1 + z')^3}}.
\end{equation} 

The energy loss due to redshifting is given by:
\begin{equation}
-(\partial \log E/\partial t)_{redshift} = H_{0}\sqrt{\Omega_{m}(1+z)^3 +
 \Omega_{\Lambda}}.
\label{redshift} 
\end{equation}

The decay widths for the VPE process are given by Equations (\ref{Gn1}) and (\ref{Gn2}) for the cases $n = 1$ and $n = 2$, respectively, while those for neutrino splitting are given by Equations (\ref{Gsplitn1}) and (\ref{Gsplitn2}). 

\section{The Theoretical Neutrino Energy Spectrum}
\label{spec}
\vspace{-6pt}
\subsection{[d] = 4 $\cal{CPT}$ Conserving Operator Dominance}

In their seminal paper, using Equation (\ref{G}), Cohen and Glashow~\cite{co11} showed how the VPE process in the $[d] = 4$ case implied powerful constraints on LIV. They obtained an upper limit of $\delta < \cal{O}$ $(10^{-11})$ based on the initial observation of high energy neutrinos by {IceCube}. Further predictions of limits on $\delta$ with cosmological factors taken into account were then made, with the predicted spectra showing a~pileup followed by a cutoff~\cite{go12}. An upper limit of $\delta < \cal{O}$ $(10^{-18})$ was obtained~\cite{bo13} based on later {IceCube} observations~\cite{aa13}. 

Using the energy loss rate given by Equation (\ref{G}), a value for $\delta_{\nu e} < 10^{-20}$ was obtained based on a model of the redshift evolution of neutrino sources and using Monte Carlo techniques to take account of propagation effects as discussed in Section \ref{cos}~\cite{st14b,st15}.~The upper limit on $\delta_{e}$ is given by $\delta_{e} \le 5 \times 10^{-21}$~\cite{st14b}. Taking this into account, one gets the constraint $\delta_{\nu} \le (0.5 - 1) \times 10^{-20}$. The~spectra derived therein for the $[d] = 4$ case also showed a pileup followed by a cutoff. The predicted cutoff is determined by redshifting the threshold energy effect. 

\subsection{[d] = 6 $\cal{CPT}$ Conserving Operator Dominance} 

In both the $[d] = 4$ and $[d] = 6$ cases, the best fit matching the theoretical propagated neutrino spectrum, normalized to an energy flux of $E_{\nu}^2(dN_{\nu}/dE_{\nu}) \simeq 10^{-8}~ {\rm GeV}{\rm cm}^{-2}{\rm s}^{-1}{\rm sr}^{-1}$ below 0.3 PeV, with the {IceCube} data corresponds to a VPE rest-frame threshold energy $E_{\nu, \rm th} = 10$ PeV, as shown in Figure~\ref{combined}~\cite{st14b,st15}.~This corresponds to $\delta_{\nu e} \equiv \delta_{\nu} - \delta_e \le \ 5.2 \times 10^{-21}$.~Given that $\delta_e \le \ 5 \times 10^{-21}$, it~is~again found that $\delta_{\nu} \le (0.5 - 1) \times 10^{-20}$. As shown in Figure \ref{thresholdeffects}, values of $E_{\nu, \rm th}$ less than 10 PeV are inconsistent with the {IceCube} data. The result for a 10-PeV rest-frame threshold energy is just consistent with the {IceCube} results, giving a cutoff effect above 2 PeV. 

In the case of the $\cal{CPT}$ conserving $[d] = 6$ operator (\emph{n} = 2) dominance, as in the $[d] = 4$ case, the~results shown in Figure~\ref{combined} show a high-energy drop off in the propagated neutrino spectrum near the redshifted VPE threshold energy and a pileup in the spectrum below that energy. This predicted drop off may be a possible explanation for the lack of observed neutrinos above 2 PeV~\cite{st14b,st15}. This~pileup is caused by the propagation of the higher energy neutrinos in energy space down to energies within a~factor of $\sim$5 below the VPE threshold. 

The pileup effect caused by the neutrino-splitting process is more pronounced than that caused by the VPE process because neutrino splitting produces two new lower energy neutrinos per interaction. This would be a potential way of distinguishing a dominance of $[d] > 4$ Planck-mass suppressed interactions from $[d] = 4$ interactions.~Thus, with better statistics in the energy range above 100 TeV, a~significant pileup effect would be a signal of Planck-scale physics. Pileup features are indicative of the fact that fractional energy loss from the last allowed neutrino decay before the VPE process ceases is 78\%~\cite{co11}, and that for neutrino splitting is taken to be 1/3. The pileup effect is similar to that of energy propagation for ultrahigh energy protons near the GZK threshold~\cite{st89}.

\begin{figure}
\centerline{\includegraphics[width=4in]{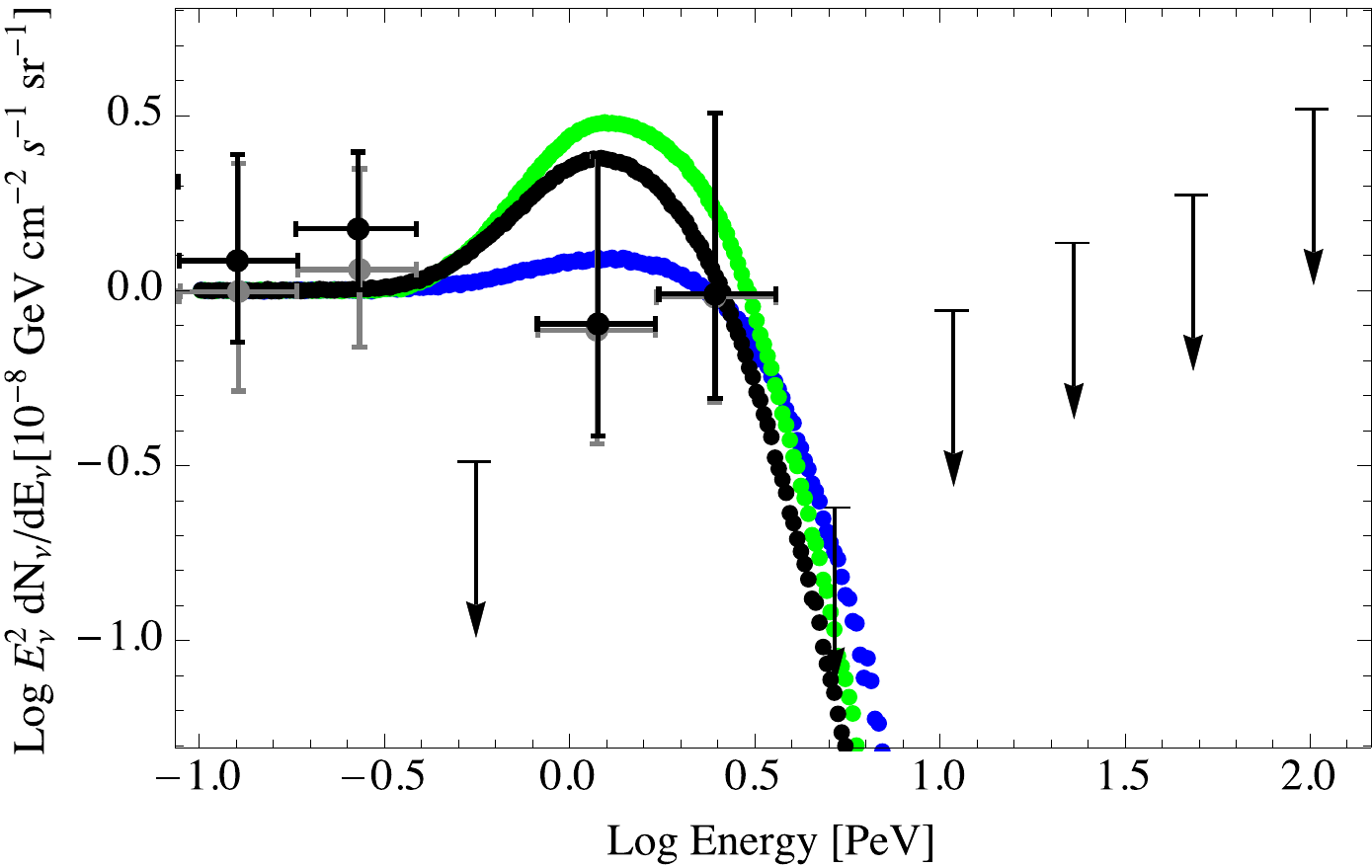}}
\caption{Propagated neutrino spectra including energy losses as described in the text\protect~\cite{st15}. Separately calculated \emph{n} = 2 neutrino spectra with the VPE case shown in blue and the neutrino splitting case shown in green. The black spectrum takes account of all three processes
(redshifting, neutrino splitting and VPE) occurring simultaneously. The rates for all cases are fixed by setting the rest frame threshold energy for VPE at 10 PeV. The neutrino spectra are normalized to the {IceCube} data both with (gray) and without (black) an estimated flux of prompt atmospheric neutrinos subtracted~\cite{aa14}.}
\label{combined}
\end{figure}

\begin{figure}
\centerline{\includegraphics[width=4.3in]{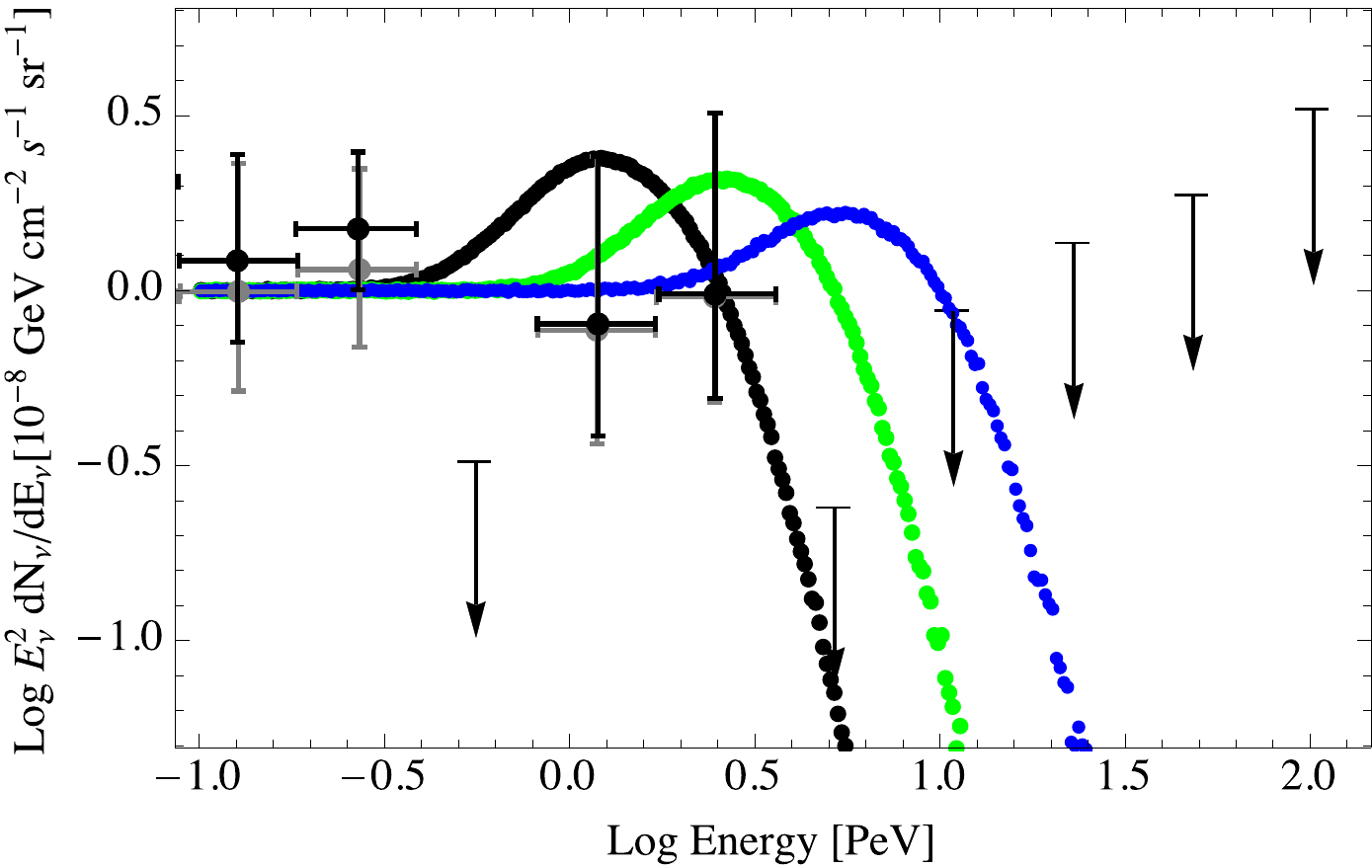}}
\caption{Calculated \emph{n} = 2 spectra taking into account all three processes
(redshifting, neutrino splitting and VPE) occurring simultaneously for rest frame VPE threshold energies of 10 PeV (black,~as~in Figure \ref{combined}), 20 PeV (green) and 40 PeV (blue). The {IceCube} data
are as in Figure~\ref{combined}~\cite{aa14}.}
\label{thresholdeffects}
\end{figure}

\subsection{[d] = 5 \cal{CPT} Violating Operator Dominance}

In the \emph{n} = 1 case, the dominant $[d] = 5$ operator violates $\cal{CPT}$.~Thus, if the $\nu$ is superluminal, the $\bar{\nu}$ will be subluminal, and {vice
versa}. However, the {IceCube} detector cannot distinguish neutrinos from antineutrinos. 
The incoming $\nu (\bar{\nu}$) generates a shower in the detector, allowing
a measurement of its energy and direction. Even in cases where there is a muon
track, the charge of the muon is not~determined. 

There would be an exception for electron antineutrinos at 6.3 PeV, 
given an expected enhancement in the event rate at the $W^{-}$ Glashow 
resonance since this resonance only occurs with $\bar{\nu_{e}}$. We note that $\nu - \bar{\nu}$ oscillation measurements would give
the strongest constraints on the difference in $\delta$'s between $\nu$'s and 
$\bar{\nu}$'s~\cite{ab15}. 

Since both VPE and neutrino-splitting interactions generate
a particle-antiparticle lepton pair, one of the pair particles will be
superluminal ($\delta > 0$), whereas the other particle will be subluminal 
($\delta < 0$)~\cite{km13}. Thus, of the daughter particles,
one will be superluminal and interact, while the other will only redshift.
The overall result in the $[d] = 5$ case is that no clear spectral cutoff occurs~\cite{st15}.

\section{Summary: Results for Superluminal Neutrinos}
\label{sum}

In the SME EFT formalism~\cite{ko89,ck98,km13}, if the apparent cutoff above $\sim 2$ PeV in the neutrino spectrum shown in Figure \ref{thresholdeffects} is caused by LIV, this would result from an EFT with either a dominant $[d] = 4$ term with $\mathaccent'27 c^{(4)} = -\delta_{\nu e} = 5.2 \times 10^{-21}$, or by a dominant $[d] = 6$ term with $\mathaccent'27 c^{(6)} = -\lambda_2/M_{Pl}^2 \ge - 5.2 \times 10^{-35}$ GeV$^{-2}$~\cite{st15}. Such a cutoff would not occur if the dominant LIV term is a $\cal{CPT}$-violating the $[d] = 5$ operator (However, see Footnote 2.)~An event with energy significantly above 3 PeV would increase our constraints by as much as an order of magnitude. However, there is still the lack of expected Glashow resonance events to consider.

If the lack of neutrinos at the Glashow resonance is the result of LIV effects as shown in Figures~\ref{combined} and \ref{thresholdeffects}, this would imply that there will be no cosmogenic~\cite{be69,st73} ultrahigh energy neutrinos. A~less~drastic effect in the cosmogenic neutrino spectrum can be caused by a violation of LIV in the hadronic sector at the level of $10^{-22}$~\cite{ss11}. 

A cutoff can naturally occur if it is produced by a maximum acceleration energy in the sources. In~that case, the parameters given above would be reduced to upper limits. However, the detection of a pronounced pileup just below the cutoff would be {prima facie} evidence of a $\cal{CPT}$-even LIV effect, possibly related to Planck-scale physics. In fact, $\cal{CPT}$-even LIV in the gravitational sector at energies below the Planck energy has been considered in the context of Ho\v{r}ava--Lifshitz gravity~\cite{ho09,po12}. 

\section{Stable Pions from LIV}
\label{pi}

Almost all neutrinos are produced by pion decay. It has been suggested that if LIV effects can prevent the decay $\pi \rightarrow \mu + \nu$ of charged pions above a threshold energy, thereby eliminating higher energy neutrinos at the Glashow resonance energy and above~\cite{an14,to15}. In order for the pion to be stable above a critical energy $E_{c}$, we require that its effective mass as given by Equation (\ref{effectivemass}) is less than the effective mass of the muon that it would decay to, i.e., $\tilde{m}_{\pi} < \tilde{m}_{\mu}$. This situation requires the condition that $\delta_{\pi} < \delta_{\mu}$~\cite{co99}. Then, neglecting the neutrino mass, this critical energy energy is given by:
\begin{equation}
E_{c} = {{\sqrt{m_{\pi}^2 - m_{\mu}^2}}\over{\delta_{\mu} - \delta_{\pi}}} 
\end{equation}
\noindent noting that if, as before, we write $\delta_{\mu\pi} \equiv \delta_{\mu} - \delta_{\pi}$; then, for small $\delta_{\mu\pi}$, it follows that $\sqrt{2\delta_{\mu\pi}} \simeq \delta_{\mu\pi}$. In~terms of SME formalism with Planck-mass suppressed terms, $\delta_{\mu\pi}$ is given by Equation (\ref{d}) or Equation (\ref{d}). For example,
if we set $E_{c}$ = 6 PeV, in order to just avoid the Glashow resonance, we get the requirement,
$ \delta_{\mu\pi} \simeq 1.5 \times 10^{-8}$. 

If a lack of multi-PeV neutrinos is caused by this effect, there will be no pileup below the cutoff energy as opposed to the superluminal cases previously discussed.~This would make the stable pion case difficult to distinguish from a natural cutoff caused by maximum cosmic-ray acceleration energies in the neutrino sources.

\section{Conclusions}

One of the most profound problems of modern physics is how to reconcile general relativity with quantum physics. A unification of these fundamental principles must somehow occur at the Planck scale. While direct observational evidence on the physics at the Planck scale itself is unattainable, within the EFT framework, clues to ``quantum gravity'' phenomena can be searched for at energies much lower than $M_{Pl}$. One of these clues may be a tiny violation of Lorentz symmetry that may increase with energy. In this regard, it has been recognized that since cosmic $\gamma$-rays and neutrinos and cosmic-ray nucleons provide the highest observable energies in the universe, high energy astrophysics observations can play an important role in searching for clues to Planck-scale physics. 

As discussed in this paper, such astrophysical searches have only thus far yielded strong constraints on LIV, verifying the strength of Lorentz symmetry.~However, there may be a hint at possible Lorentz symmetry breaking at the level of $10^{-20}$ in the neutrino sector at PeV energies. Better~statistics are needed to explore this possibility further.

\vspace{6pt}
Acknowledgments: I thank my collaborators in much of the work discussed here: Stefano Liberati, David~Mattingly and Sean Scully.



\begin{thebibliography}{999}
\bibitem{pl06}
Planck, M. \emph{The Theory of Radiation}; Translated from 1906; Dover Publications: New York, NY, USA, 1959.

\bibitem{ko89} Kosteleck\'{y}, V.A.; Samuel, S. Spontaneous breaking of Lorentz symmetry in string theory. 
\emph{Phys. Rev. D} {\bf 1989}, \emph{39}, 683-685. 

\bibitem{ta14} Tasson, J.D. What do we know about Lorentz invariance? \emph{Rpt. Prog. Phys.} {\bf 2014}, \emph{77}, 062901.

\bibitem{he16} Hees, A. et al. Tests of Lorentz Symmetry in the Gravitational Sector. \ %
\emph{Universe} {\bf 2016}, \emph{2}, 30.

\bibitem{co99} Coleman, S.R.; Glashow, S.L. High-energy tests of Lorentz invariance. \emph{Phys. Rev. D} {\bf 1999}, \emph{59}, 116008.

\bibitem{sg01} Stecker, F.W.; Glashow, S.L. New tests of Lorentz invariance following from observations of the highest energy cosmic $\gamma$-rays. \emph{Astropart. Phys.} {\bf 2001}, \emph{16}, 97-99.

\bibitem{st17} Stecker, F.W. Search for the footprints of new physics with laboratory and cosmic neutrinos. \emph{Mod. Phys. Lett. A} \textbf{2017}, {\it 32}, 1730014, arXiv:1705.08485.

\bibitem{ta98} Tanimori, T.; et al. Detection of $\gamma$-rays up to 50 TeV from the Crab Nebula. \emph{Astrophys. J.} {\bf 1998}, \emph{492}, L33-L36.

\bibitem{ni80} Nishimura, J.; et al. Emulsion chamber observations of primary cosmic-ray electrons in the energy range 30-1000 GeV. \emph{Astrophys. J.} {\bf 1980}, \emph{238}, 394-409.

\bibitem{ah01} Aharonian, F.A.; et al. Reanalysis of the high energy cutoff of the 1997 Mkn 501 TeV energy spectrum. \emph{Astron. Astrophys. }{\bf 2001}, \emph{366}, 62-67.

\bibitem{st92} Stecker, F.W.; De Jager, O.C.; Salamon, M.H. TeV gamma rays from 3C 279 - A possible probe of origin and intergalactic infrared radiation fields. \emph{Astrophys. J.} {\bf 1992}, {\it 390}, L49-L52.

\bibitem{ds02} de Jager, O.C.; Stecker, F.W. Extragalactic Gamma-Ray Absorption and the Intrinsic Spectrum of Markarian 501 during the 1997 Flare. \emph{Astrophys. J.} {\bf 2002}, \emph{566}, 738-743.

\bibitem{ko03} Konopelko, A.; et al. Modeling the TeV Gamma-Ray Spectra of Two Low-Redshift Active Galactic Nuclei: Markarian 501 and Markarian 421.{\it Astrophys. J.} {\bf 2003}, {\it
597}, 851-859.

\bibitem{ja08a} Jacob, U.; Piran, T. Inspecting absorption in the spectra of extra-galactic gamma-ray sources for insight into Lorentz invariance violation. \emph{Phys. Rev. D} {\bf 2008}, \emph{78}, 124010.

\bibitem{gr66} Greisen, K. End to the cosmic-Ray spectrum? 
{\it Phys. Rev. Lett.} {\bf 1966}, \emph{16}, 748-750.

\bibitem{za66} Zatsepin, G.T.; Kuz'min V.A. Upper Limit of the spectrum of cosmic rays.  
\emph{Zh. Eks. Teor. Fiz. Pis'ma Red.} Kuz'min, V. A.  {\bf 1966}, \emph{4}, 144-146

\bibitem{st68} Stecker, F.W. Effect of photomeson production by the universal radiation field on high-energy cosmic rays. {\it Phys. Rev. Lett.} {\bf 1968}, \emph{21}, 1016-1018.

\bibitem{ss09} Scully, S.T.; Stecker, F.W. Lorentz invariance violation and the observed spectrum of ultrahigh energy cosmic rays. {\it Astropart. Phys.}
{\bf 2009}, \emph{31}, 220-225.

\bibitem{al03} Alfaro, J.; Palma, G. Loop quantum gravity and ultrahigh energy cosmic rays. {\it Phys. Rev. D} {\bf 2003}, \emph{67}, 083003.

\bibitem{sch09} Sch\"ussler, F. (Auger Collaboration.) The Cosmic Ray Energy Spectrum and Related Measurements with the Pierre Auger Observatory. {\it Proceedings of the 31st International Cosmic Ray Conference}, \L\'od\'z, Poland, 7--15 July 2009.

\bibitem{ss11} Scully, S.T.; Stecker, F.W. Testing Lorentz invariance with neutrinos from ultrahigh energy cosmic ray interactions. \emph{Astropart. Phys.} {\bf 2011}, \emph{34}, 575-580.

\bibitem{al16} Allison, P.; et al. Performance of two Askaryan Radio Array stations and first results in the search for ultrahigh energy neutrinos. \emph{Phys. Rev. D} {\bf 2016}, \emph{93}, 082003.

\bibitem{ba17} Barwick, S.; et al. Radio detection of air showers with the ARIANNA experiment on the Ross Ice Shelf. \emph{Astropart. Phys.}
{\bf 2017}, \emph{90}, 50-68.

\bibitem{st04} Stecker, F.W.; et al.; Observing the ultrahigh energy universe with OWL eyes.\emph{Nucl. Phys. B} {\bf 2004}, \emph{136C}, 433-438.

\bibitem{ol17} Olinto, A. POEMMA: Probe Of extreme multi-messenger astrophysics.  {\it Proceedings of the 35th International Cosmic Ray Conference}, Busan, Korea, 12--20 July 2017.

\bibitem{ja08} Jacob, U.; Piran, T. Lorentz-violation-induced arrival delays of cosmological particles. \emph{J. Cosmol. Astropart. Phys.} {\bf 2008}, \emph{1}, 031.

\bibitem{vv13} Vasilieou, V.; et al. Constraints on Lorentz invariance violation from Fermi-Large Area Telescope observations of gamma-ray bursts. \emph{Phys. Rev. D} {\bf 2013}, \emph{87}, 122001.

\bibitem{ac98} Amelino-Camelia, G.; et al. Tests of quantum gravity from observations of $gamma$-ray bursts. \emph{ Nature} {\bf 1998}, \emph{393}, 763-765.

\bibitem{el08} Ellis, J.; Mavromatos, N.E.; Nanopoulos, D.V. Vacuum refractive index. 
\emph{Phys. Lett. B} {\bf 2008}, \emph{665}, 412-417.

\bibitem{ki99} Kifune, T. Invariance violation extends the cosmic-ray horizon? \emph{Astrophys. J. Lett.} {\bf 1999}, {\it 518}, L21-L24.

\bibitem{mp03} Myers, R.C.; Pospelov, M. Pospelov, Ultraviolet modifications of dispersion relations in effective field theory. \emph{Phys. Rev. Lett.} {\bf 2003}, \emph{90}, 211601.

\bibitem{ck98} Colladay, D.; Kostelecky, V.A. Lorentz-violating extension of the standard model.
\emph{Phys. Rev. D} {\bf 1998}, \emph{58}, 116002.

\bibitem{km13} Kosteleck\'{y}, V.A.; Mewes, M. Fermions with Lorentz-violating operators of arbitrary dimension. \emph{Phys. Rev. D} {\bf 2013}, \emph{88}, 096006.

\bibitem{st11} Stecker, F.W. A new limit on Planck scale Lorentz violation from $\gamma$-ray burst polarization. \emph{Astropart. Phys.} {\bf 2011},\emph{ 35}, 95-97.

\bibitem{go14} G\"{o}tz, D.; et al. GRB 140206A: The most distant polarized gamma-ray burst \emph{Mon. Not. R. Astron. Soc.} {\bf 2014}, \emph{444}, 2776-2782.

\bibitem{st15} Stecker, F.W.; Scully, S.; Liberati, S.; Mattingly, D. Searching for traces of Planck-scale physics with high energy neutrinos. \emph{Phys. Rev. D} {\bf 2015}, \emph{91}, 045009.

\bibitem{ko12} Kosteleck\'{y}, V.A.; Mewes, M. Neutrinos with Lorentz-violating operators of arbitrary dimension. \emph{Phys. Rev. D} {\bf 2012}, \emph{85}, 096005.

\bibitem{li13} S. Liberati, Tests of Lorentz invariance: a 2013 update. \emph{Classical Quantum Gravity} {\bf 2013} \emph{30}, 133001.

\bibitem{gr02} Greenberg, O.W. CPT Violation Implies Violation of Lorentz Invariance. \emph{Phys. Rev. Lett.} \textbf{2002}, \emph{89}
231602.

\bibitem{ko09} Kosteleck\'{y}, V.A.; Mewes, M. Electrodynamics with Lorentz-violating operators of arbitrary dimension. \emph{Phys. Rev. D} {\bf 2009}, \emph{80}, 015020.

\bibitem{st14b} Stecker, F.W. Limiting superluminal electron and neutrino velocities using the 2010 Crab Nebula flare and the IceCube PeV neutrino events.
\emph{Astropart. Phys.} {\bf 2014}, \emph{56}, 16-18.

\bibitem{co11} Cohen, A.G.; Glashow, S.L. Pair creation constrains superluminal neutrino propagation. \emph{Phys. Rev. Lett.} {\bf 2011}, \emph{107}, 181803.

\bibitem{ca12} Carmona, J.M.; Cort\'{e}s, J.L.; Maz\'{o}n, D. Uncertainties in constraints from pair production on superluminal neutrinos. \emph{Phys. Rev. D} {\bf 2012}, \emph{85}, 113001.

\bibitem{gg04} Gonzalez-Garcia, M.C.; Maltoni, M. Atmospheric neutrino oscillations and new physics. \emph{Phys. Rev. D} {\bf 2004}, \emph{70}, 033010.

\bibitem{ab15} Abe, K.; et al. (Super-Kamiokande.)
Test of Lorentz invariance with atmospheric neutrinos.
\emph{Phys. Rev. D} {\bf 2015}, \emph{91}, 052003.

\bibitem{gg16} Gonzalez-Garcia, M.C.; Maltoni, M. Global analyses of neutrino oscillation experiments. \emph{Nucl. Phys. B} {\bf 2016}, \emph{908}, 199-217.

\bibitem{mac13} Maccione, L.; Liberati, S.; Mattingly, D. Violations of Lorentz invariance in the neutrino sector: an improved analysis of anomalous threshold constraints. \emph{JCAP} {\bf 2013}, \emph{3}, 039.

\bibitem{aa15} Aartsen, M.G.; et al. (IceCube) Atmospheric and astrophysical neutrinos above 1 TeV interacting in IceCube. \emph{Phys. Rev. D} {\bf 2015}, \emph{91}, 022001.

\bibitem{aa14} Aartsen, M.G.; et al.(IceCube) Observation of high-energy astrophysical neutrinos in three years of IceCube data. \emph{Phys. Rev. Lett.} {\bf 2014}, \emph{113}, 101101.

\bibitem{aa13} Aartsen, M.G.; et al. (IceCube) The IceCube observatory at the South Pole detected neutrinos from outside our solar system. \emph{Science} {\bf 2013}, \emph{342}, 1242856.

\bibitem{st79} Stecker, F.W. Diffuse fluxes of cosmic high-energy neutrinos. \emph{Astrophys. J.} {\bf 1979}, \emph{228}, 919-927.

\bibitem{ah14} Ahlers, M.; Murase, K. Probing the galactic origin of the IceCube excess with gamma-rays. \emph{Phys. Rev. D} {\bf 2014}, \emph{90}, 023010.

\bibitem{gl60} Glashow, S.L. Resonant Scattering of Antineutrinos \emph{Phys. Rev.} {\bf 1960}, \emph{118}, 316-317.

\bibitem{aa16b} Aartsen, M.G.; et al. (IceCube) Observation and Characterization of a Cosmic Muon Neutrino Flux from the Northern Hemisphere Using Six Years of IceCube Data.  \emph{Astrophys. J.} {\bf 2016}, \emph{833}, 3.
 

\bibitem{be13} Behroozi, P.S.; Wechsler, R.H.; Conroy, C. The Average Star Formation Histories of Galaxies in Dark Matter Halos from z = 0-8. \emph{Astrophys. J.} {\bf 2013}, \emph{770}:57.

\bibitem{go12} Gorham, P.W.; et al. Implications of ultrahigh energy neutrino flux constraints for Lorentz-invariance violating cosmogenic neutrinos. \emph{Phys. Rev. D} {\bf 2012}, \emph{86}, 103006.

\bibitem{bo13} Borriello, E.; Chakraborty, S.; Mirizzi, A. Stringent constraint on neutrino Lorentz invariance violation from the two IceCube PeV neutrinos. \emph{Phys. Rev. D} {\bf 87}, 116009.

\bibitem{st89} Stecker, F.W. Extragalactic radiation and the ultra-high-energy cosmic-ray spectrum. \emph{Nature} {\bf 1989}, \emph{342}, 401-403.

\bibitem{be69} Berezinsky, V.; Zatsepin, G.T. Cosmic rays at ultrahigh energy (neutrino?). {\it Phys. Lett. B} {\bf 1969}, \emph{28}, 423-424.

\bibitem{st73} Stecker, F.W. Ultrahigh energy photons, electrons, and neutrinos, the microwave background, and the universal cosmic-ray hypothesis. \emph{Ap. Sp. Sci.} {\bf 1973}, \emph{20}, 47-57. 

\bibitem{ho09} Ho\v{r}ava, P. Quantum gravity at a Lifshitz point. \emph{Phys. Rev. D} {\bf 2009}, \emph{79}, 084008.

\bibitem{po12} Pospelov, M.; Shang, Y. Lorentz violation in Ho?ava-Lifshitz-type theories.\emph{Phys. Rev. D} {\bf 2012}, \emph{85}, 105001.

\bibitem{an14} Anchordoqui, L.A.; et al. End of the cosmic neutrino energy spectrum. \emph{Phys. Lett. B} {\bf 2014}, \emph{739}, 99-101.

\bibitem{to15} Tomar, G.; Mohanty, S.; Pakvasa, S. Lorentz invariance violation and IceCube neutrino events. \emph{J. High Energy Phys.} {\bf 2015}, \emph{11}, 022.








\end{thebibliography}


\end{document}